\documentclass[preprint,preprintnumbers,amsmath,amssymb]{revtex4}

\usepackage{epsfig}
\usepackage{graphicx}
\usepackage{dcolumn}
\usepackage{bm}

\begin{document}

\renewcommand{\baselinestretch}{1.5}

\renewcommand{\thefootnote}{\fnsymbol{footnote}}

\title{Tunneling magnetoresistance in devices based on epitaxial NiMnSb with uniaxial anisotropy}

\author{J. Liu*}
\author{E. Girgis} \author{P. Bach} \author{C. R\"{u}ster}
\author{C. Gould} \author{G. Schmidt} \author{L.W. Molenkamp}

\affiliation{Physikalisches Institut der Universit\"{a}t W\"{u}rzburg, Am
Hubland, 97074 W\"{u}rzburg, Germany}

\affiliation{*Engineering Research Center for Semiconductor Integrated
Technology, Institute of Semiconductors, Chinese Academy of Sciences, P.O. Box
912, Beijing 100083, P.R. China}

\date{\today}

\begin{abstract}
We demonstrate tunnel magnetoresistance (TMR) junctions based on a tri layer
system consisting of an epitaxial NiMnSb, aluminum oxide and CoFe tri layer.
The junctions show a tunnelling magnetoresistance of $\Delta R/R$ of 8.7\% at
room temperature which increases to 14.7\% at 4.2K. The layers show clear
separate switching and a small ferromagnetic coupling. A uniaxial in plane
anisotropy in the NiMnSb layer leads to different switching characteristics
depending on the direction in which the magnetic field is applied, an effect
which can be used for sensor applications.


\end{abstract}

\maketitle

\section{Introduction}
Tunneling magnetoresistance (TMR) was first observed in
ferromagnet-insulator-ferromagnet (FM-I-FM) junctions using ferromagnetic
metals (like CoFe, Co or Fe) and Al$_2$O$_3$ tunnel
barriers\cite{Moodera95,Miyazaki95}. Thereafter TMR has been intensively
studied due to its applications in magnetic random-access-memory (MRAM) and
magnetic sensors\cite{Moodera99,Wolf01}. The magnitude of the tunneling
magnetoresistance depends strongly on the density of states in the electrodes
and in most cases  can be calculated using Jullieres formula\cite{Julliere75}.

\begin{equation}
TMR=\frac{R_{AP} - R_P}{R_{AP}}=\frac{2P_1 P_2}{1+P_1 P_2} \label{equa1}
\end{equation}

where P$_1$ and P$_2$ are the spin polarization for the two FMs. From this formula it can
immediately be inferred that a higher spin polarization in the electrodes also leads to
an increase of the TMR. For a material in which all tunnelling electrons have the same
spin direction, the magnetoresistance can theoretically be infinite. This fact has led to
an increasing interest in the possibility of using half metallic materials for TMR
junctions and indeed experiments with $La_{0.7}Sr_{0.3}MnO_3$ \cite{LSMO} have already
yielded extremely high TMR values.

Other materials for which half metallic behaviour is predicted can be found
among the so called Heusler alloys\cite{Heusler1903}, which are typically
ternary compounds with metallic conductivity. One candidate which has been
subject to several theoretical calculations is NiMnSb which is predicted to be
100\% spin polarized even at room temperature\cite{Groot83}. Because of this
theoretical prediction the material and its deposition have been studied by
various groups and it is now possible to obtain fully epitaxial thin films of
NiMnSb\cite{Vanroy00,Bach03}. Positron annihilation experiments on bulk
material have already indicated that NiMnSb is indeed half
metallic\cite{Hanssen86,Hanssen90}. Despite the high quality of epitaxial
layers, however, no half-metallicity in thin layers has been observed up to
now. A possible explanation is that most investigation methods for layers are
surface sensitive and surface states may cause the appearence of additional
states in the gap of the bandstructure of the NiMnSb, leading to a reduced spin
polarization\cite{Fang02}.

However, even with a spin polarization of slightly less than 100 \% NiMnSb can
still be interesting for TMR devices and a number of attempts have been made to
introduce single or polycristalline NiMnSb layers in various kinds of
structures\cite{Tanaka97}. The magnetoresistance obtained up to now, however is
comparable to or even smaller than that achieved using traditional metallic
ferromagnets.

We have fabricated a layer stack based on a 2.5 (???) nm thick epitaxial NiMnSb
layer grown on a lattice matched semiconductor substrate. This layer has a very
low switching field and, due to its thinness\cite{Bach03} a strong in
plane-uniaxial anisotropy. $Al_2O_3$ is used as a tunnel barrier and CoFe as
the counter electrode. In this paper, we present the fabrication of the
devices, along with the results of magnetoresistance measurements for various
field directions and temperatures, and the influence of the anisotropy on the
magnetoresistance is explained.

The NiMnSb layer is deposited in a multi-chamber molecular beam epitaxy (MBE)
system. An InP substrate with a lattice constant of $5.869\AA$ close to that of
NiMnSb ($5.903\AA$) allows for the growth of tens of nm of fully strained
NiMnSb\cite{Bach03}. In a first step, a 300 nm thick highly n-doped (In,Ga)As
buffer is deposited on the InP. The sample is subsequently transferred into the
NiMnSb growth chamber, where the NiMnSb layer is deposited at a substrate
temperature of 300 °C. (Detailed growth conditions can be found in ref.
\cite{Bach03}.) Subsequently, the sample is transferred into a UHV sputtering
chamber where a 0.9 nm thick Al layer is deposited by magnetron sputtering. In
a separate chamber, this layer is then oxidized at room temperature in pure
oxygen with a pressure of $10^4$ Pa. A second layer of Al is deposited and
again oxidized. With this process, a tunnel barrier with a resistance of
approx. 150 $k\Omega$cm is obtained. Finally a 10 nm layer of CoFe is deposited
on the AlOx and the sample is covered by a 2 nm thick Pt layer in order to
prevent oxidation.

Mesa structures are fabricated by optical lithography and Argon ion-beam
etching down to the highly conductive (In,Ga)As layer. The mesas are passivated
by PECVD deposition of a layer of $Si_3N_4$. Windows are opened on the top of
the Mesa (top contact) and on the (In,Ga)As close to the Mesa (bottom contact).
Using optical lithography, deposition of Ti/Au, and lift off, two large pads
are created, which contact top and bottom of the structure through the windows
in the $Si_3N_4$ and allow for bonding far away from the mesa without damaging
the TMR contact.

The structures are characterized using a DC transport measurement setup, an
electromagnet for room temperature measurements and a superconducting magnet in
a $^4$He bath cryostate for low temperatures. A Quantum Design MPMS SQUID
magnetometer is used to determine the switching properties of the magnetic
layers at both room temperature and 5 K.

The SQUID measurements are done on samples about 3x3 mm in size. The magnetic
field is applied in plane and along the easy axis of the NiMnSb layer [110].
The sample shows a clear separate switching of the two magnetic layers both at
RT and at 5 K. The CoFe layer shows a coercive field of 115 Oe at room
temperature and 126 Oe at 5 K the coercive field, consistent with expectations.
For the NiMnSb layer the coercive field is lower than for the CoFe, however, it
is much higher than has been previously observed for single NiMnSb layers. This
suggests that some ferromagnetic coupling is present between the two layers.
This coupling, which evidently increases at low temperatures, combined with the
large magnetic moment of the CoFe layer is responsible for the high coercive
field.

\section{Results and discussions}
\begin{figure}
\begin{center}
\resizebox{10 cm}{8 cm}{\includegraphics{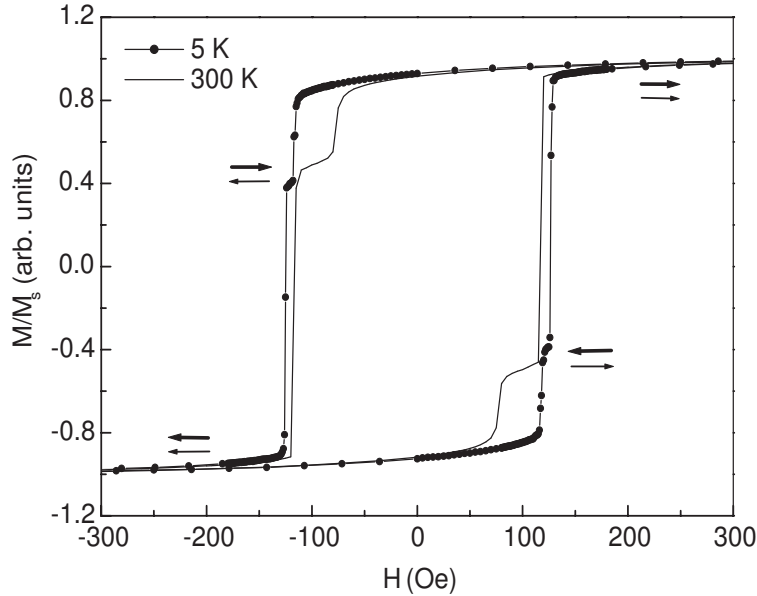}} \caption{SQUID
magnetization measurements of the TMR stack at 4.2 and 300~K. The signal shows
clear separate switching of the two layers.} \label{squid}
\end{center}\end{figure}

I/V measurements on the samples at room temperature and at 4.2 K indicate the presence of
a tunnel barrier. The shape of the dI/dV curve can be fitted to Simmon's
formula\cite{Simmons63} yielding a barrier height of 1.65 eV and a barrier thickness of
1.9 nm which is in good agreement with the thickness of the deposited Al.

\begin{figure}
\begin{center}
\resizebox{11 cm}{8.8 cm}{\includegraphics{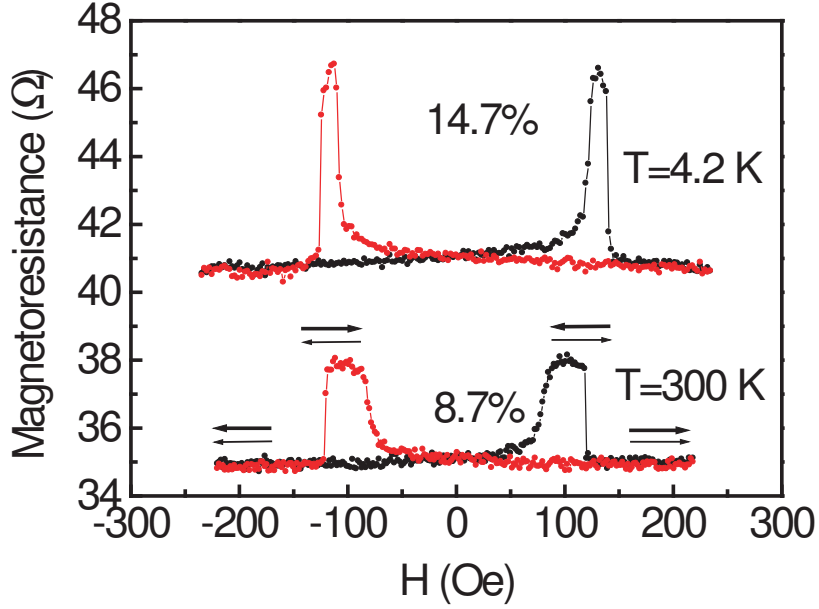}} \caption{Magnetoresistance
measurements with \textbf{H} along the easy-axis of the NiMnSb layer at 4.2 and 300~K.
The arrows show the magnetization direction for the NiMnSb (thin arrow) and CoFe layers
(thick arrow).} \label{TMR}
\end{center}\end{figure}

\begin{figure}
\begin{center}
\resizebox{10 cm}{8 cm}{\includegraphics{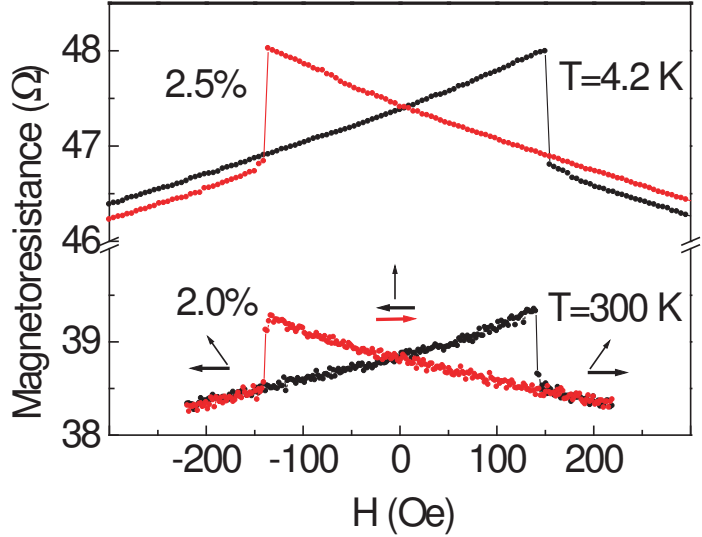}} \caption{Magnetoresistance
measurements with \textbf{H} along the hard-axis of the NiMnSb layer at 4.2 and 300~K,
respectively. The arrows show the magnetization direction for the NiMnSb (thin arrow) and
CoFe layers (thick arrow). The red arrow shows the orientation of the CoFe layer for the
downscan} \label{TMR_hard}
\end{center}\end{figure}

Fig. \ref{TMR} shows magnetoresistance measurements done with the magnetic field applied
along the easy axis of the NiMnSb layer. At room temperature as well as at 4.2 K the
magnetoresistance traces have the typical shape expected from a TMR device with
independent switching of the two magnetic layers. Only from the high switching field of
the NiMnSb and the narrowing of the high resistance region at low temperatures can a
ferromagnetic coupling be inferred. The TMR ratio $\Delta R/R$ is 8\% at room temperature
and 14\% at 4.2 K. When the magnetic field {\bf\textit{H}} is applied along the in-plane
hard axis of the NiMnSb as in fig. \ref{TMR_hard}, the shape of the curve is changed.
Starting from negative saturation we see a linear increase of the resistance which
persists until {\bf\textit{H}} has crossed zero. Then at moderate fields the resistance
drops by roughly 2\% in a single step and finally drops linearly to the initial
saturation value. The saturation field at 4.2 K is as high as 200 mT.

The shape and magnitude of the MR traces for the easy axis measurements are easily
explained by taking into account collinear magnetization for the NiMnSb and the CoFe. At
large negative fields we have parallel alignment which changes to antiparallel alignment
at small positive {\bf\textit{H}} when the NiMnSb layer reverses its magnetization and
the high resistance (antiparallel) state is obtained (See arrows in fig \ref{TMR}).
Further increase of {\bf\textit{H}} leads to the reversal of the CoFe layer and again to
a low resistance (parallel) state. For the hard axis measurements we also start with
parallel alignment for large negative fields. The crystalline anisotropy of the NiMnSb,
however, already starts to dominate at negative fields and the magnetization vector of
the NiMnSb layer begins to rotate towards its easy axis which is perpendicular to the
magnetic field. This rotation leads to an increase in resistance because the
magnetization of the CoFe and the NiMnSb are no longer collinear. At zero
{\bf\textit{H}}-field the magnetizations are perpendicular to each other. When
{\bf\textit{H}} is increased beyond zero the magnetization of the NiMnSb slightly rotates
towards the antiparallel state and the resistance further increases. At the coercive
field of the CoFe layer the resistance exhibits a step down when the magnetization of the
CoFe reverses. This step, however, is relatively small because the final state is still
an almost perpendicular configuration with only a slightly different projection of the
two magnetizations before and after the switching event. Further increase of the field
restores the parallel configuration by rotating the magnetization of the NiMnSb back to
the direction of the magnetic field. The fact that nowhere in this scenario an
antiparallel configuration appears explains the reduced total MR with respect to the easy
axis values.

\section{Summary}
TMR devices, which show a TMR of up to 14\% at low temperature have been fabricated based
on epitaxial NiMnSb layers. While this value is still below those reported for
conventional metallic ferromagnets, further optimization may improve the
magnetoresistance ratio. The devices presented, however, have a strong uniaxial
anisotropy which has not been observed in other NiMnSb based tunnel junctions and which
can be useful for sensor applications, especially, when an angular dependence of the
signal is required.

\begin{acknowledgments}
The authors thank V. Hock for processing. The work was supported by the German BMBF
(VDI), and the DARPA SpinS program.
\end{acknowledgments}

\end{document}